\documentclass[amsmath,amssymb,aps,superscriptaddress,reprint]{revtex4-2}

\usepackage{graphicx}
\usepackage{dcolumn}
\usepackage{bm}
\usepackage{epsf,verbatim,pifont}
\usepackage{hyperref,cleveref}
\usepackage{subfigure,tabularx}
\usepackage[dvipsnames]{xcolor}
\usepackage{kotex}
\usepackage{ulem,color}
\usepackage{xr}
\usepackage{comment}

\newcommand{\qPotts}[1]{$q$Potts}
\newcommand{\qrPotts}[1]{$(q,r)$Potts}

\usepackage{tikz}

\usepackage{xfp}
\newcommand\SupplementaryMaterials{%
    \title{Summplementary Information for \\ Entropy-Induced Phase Transitions in a Hidden Potts Model}
    \xdef\presupfigures{\arabic{figure}}
    \xdef\presupsections{\arabic{section}}
    \renewcommand\thefigure{\fpeval{\arabic{figure}-\presupfigures}}
    \renewcommand\thesection{\fpeval{\arabic{section}-\presupsections}}
}

\def\min{\textrm{min}}

\def\up{\text{up}}
\def\lo{\text{lo}}

\begin{document}

\title{Entropy-Induced Phase Transitions in a Hidden Potts Model}

\author{Cook Hyun Kim}
\affiliation{Center for Complex Systems, KI of Grid Modernization, Korea Institute of Energy Technology, Naju, Jeonnam 58217, Korea}
\author{D.-S. Lee}
\affiliation{School of Computational Sciences, and Center for AI and Natural Sciences, Korea Institute for Advanced Study, Seoul, 02455, Korea}
\author{B. Kahng}
\email{bkahng@kentech.ac.kr}
\affiliation{Center for Complex Systems, KI of Grid Modernization, Korea Institute of Energy Technology, Naju, Jeonnam 58217, Korea}
\date{\today}

\begin{abstract}
A hidden state in which a spin does not interact with any other spin contributes to the entropy of an interacting spin system. Using the Ginzburg-Landau formalism in the mean-field limit, we explore the $q$-state Potts model with extra $r$ hidden states. We analytically demonstrate that when $1 < q \le 2$, the model exhibits a rich phase diagram comprising a variety of phase transitions such as continuous, discontinuous, two types of hybrids, and two consecutive second- and first-order transitions; moreover, several characteristics such as critical, critical endpoint, and tricritical point are identified. The critical line and critical end lines merge in a singular form at the tricritical point. Those complex critical behaviors are not wholly detected in previous research because the research is implemented only numerically. We microscopically investigate the origin of the discontinuous transition; it is induced by the competition between the interaction and entropy of the system in the Ising limit, whereas by the bi-stability of the hidden spin states in the percolation limit. Finally, we discuss the potential applications of the hidden Potts model to social opinion formation with shy voters and the percolation in interdependent networks. 
\end{abstract}

\pacs{89.75.Hc, 64.60.ah, 05.10.-a}

\maketitle


\section{Introduction}
Phase transitions and critical phenomena in spin models in thermally equilibrium states~\cite{Potts, Wu, NP} can help understand collective behavior in non-equilibrium complex systems through mathematical correspondence. For instance, whereas the $q$-state Potts (denoted as Q-Potts) model widely used in statistical physics mainly to explore phase transitions in various magnetic materials can be mapped in the $q\to 1$ limit to the percolation transition model, where a giant cluster emerges as links are added in complex networks~\cite{KF_Percolation_1, KF_Percolation_2, Giri_Percolation, Wu_Percolation, Newman}. The formalism of the Bose$–$Einstein condensation in thermal quantum systems can be applied to describe the evolving structure in non-equilibrium complex networks~\cite{BE_Bianconi}. The phase transition of the Ising model emerging at critical temperatures~\cite{Ising_1967, Ising_MFT_Bianconi, Ising_MFT_SHLee, Vito_Spin, Potts-ComplexNetwork1, Potts-ComplexNetwork2, Potts-ComplexNetwork3} provides potential insights into understanding the formation of consensus in the voter model~\cite{redner, Clifford, Holley, Cox, Liggett-book1, Liggett-book2, Liggett1, Liggett2, Halu, Masuda, Diakonova, Diakonova_Latora, Chmiel1, Chmiel2, Castellano, Vieira}, despite the inherent differences between thermal and stochastic noises. Each spin represents a voter, and each spin's up or down state represents the left or right wing of the voter's opinion state.

Conversely, phase transitions in non-equilibrium systems~\cite{redner, Durrett, Neuro-Hopfield} inspire motivation to study corresponding spin models. Recently, various types of phase transitions, such as continuous and discontinuous transitions, but also hybrid and consecutive transitions have been discovered in non-equilibrium complex systems, particularly in multilayer networks. However, no universal theory has been established in the non-equilibrium systems, so underlying mechanisms are not yet fully understood. Under this circumstance, analytical methods established in the corresponding equilibrium systems can give a clue for a unified theory or framework of the non-equilibrium systems, as phase transitions and critical phenomena in many non-equilibrium systems with the Ising spin, such as the majority-vote model, belong to the same universality class as the Ising model~\cite{Oliveira_Entropy_Production}. 
Therefore, for example, understanding opinion separation or formation on two different issues in social networks can be inspired by the study of the Ashkin–Teller (denoted as AT) model on scale-free networks~\cite{AT, AT1, AT2, MS}. Its Hamiltonian comprises two types of Ising spins with intra- and inter-type interactions. The AT model's phase diagram, particularly within the mean-field limit, exhibits various types of phase transitions and critical points. A hybrid (or mixed order) phase transition occurs at a critical endpoint at which a second-order transition line intrudes into a first-order transition line. Through the complex phase diagram, the AT spin model can illustrate the causes leading to hybrid phase transitions. Moreover, the AT model in the mean-field limit can give an insight into the hybrid percolation transition in interdependent networks~\cite{mcc} where the origin cannot be explained without such an integrated scheme. Therefore, studying spin models in equilibrium systems motivated by real-world phenomena in non-equilibrium systems and vice versa is interesting and meaningful; nevertheless, this approach may not fully capture transient behaviors before a system reaches a steady state. 

With this background, this paper explores phase transitions in the Potts model with hidden states. This modified Potts model provides a rich phase diagram with various types of phase transitions and critical points, extending the model's ability to capture complex phenomena in systems beyond traditional thermodynamics. This study is motivated by the presence of so-called shy voters who do not readily express their opinions or candidate preferences until they cast their votes or forever and rarely interact with other voters~\cite{Blais_shy_voter, Gallego_shy_voter}. Consequently, opinion polls may struggle to predict the election outcome accurately. The Potts model, incorporating hidden states as defined below, can provide insights into the potential complexity of election results influenced by the presence of these shy voters.

This Potts model with hidden states was initially proposed to explain experimental results in physically disordered systems~\cite{tamura_2008, tamura_2010, Okumura, tanaka_2011, Stoudenmire}, studied recently~\cite{Krasnytska}. The hidden Potts model comprises $N$ spins, each spin taking one of $q$ visible or $r$ hidden states, $s_i=1,\cdots,q, q+1,\cdots, q+r$. Hence, it is called the $(q,r)$-state Potts model~\cite{tamura_2010, tanaka_2011, Krasnytska} and denoted as the QR-Potts model. The Hamiltonian is
\begin{equation}
\mathcal{H}=-J\sum_{\langle i,j \rangle}\sum_{\alpha=1}^{q}\delta_{s_i, \alpha} \delta_{s_j,\alpha}-H \sum_{i} \delta_{s_i, 1},
\label{eq:hamiltonian_Hidden}
\end{equation}
where $\langle i,j \rangle$ is the nearest neighbors, $\delta_{a,b}$ is the Kronecker delta function, $J >0$ is a coupling constant, and $H$ is an external field in the direction of the first state $s_i=1$. The spin $s_i$ interacts with the neighboring spins and contributes to the energy only in one of the visible states $s_i=1,2 \cdots, q$. The hidden states contribute only to the entropy. This Hamiltonian can be rewritten as 
$$ \mathcal{H}=-J\sum_{\langle i,j \rangle}\sum_{\alpha=1}^{q}\delta_{s_i, \alpha} \delta_{s_j,\alpha}-H\sum_{i} \delta_{s_i, 1}-T\ln r \sum_i \delta_{s_i,q+1}.$$
This Hamiltonian with $q=2$ was proposed to describe the two-step transition exhibited by a bi-nuclear spin-crossover complex~\cite{wajnflasz} and the solid-fluid transitions of lipidic chains~\cite{doniach}.

The mean-field (MF) solution has been obtained at $q=2$ to demonstrate a discontinuous phase transition for $r > r_c = 4e/3$~\cite{tamura_2008, tamura_2010, Okumura, tanaka_2011}. This result was surprising because the Ising model is known to exhibit only a continuous transition even in the MF limit. The authors of Ref.~\cite{Krasnytska} numerically studied the phase transitions for $1 \le q < 2$ to discover two characteristic values, $r_\ell$ and $r_h$, such that for $r_\ell < r < r_h$, as temperature $T$ decreases, the order parameter first undergoes a continuous transition and then a discontinuous transition while only a discontinuous transition occurs at $r=r_h$. Despite discovering such interesting transition behaviors, a rigorous theoretical study has been absent to elucidate the nature and origin of various phase transitions. Note that $q$ is regarded as a non-integer number, even though $q$ represents the number of states. This generalization allows us to consider the crossover behavior from the Ising limit to the percolation limit. Moreover, the Potts model with non-integer $q$ was considered in association with the cluster-weighted percolations~\cite{Sweeny_WCP, Larsson_WCP}.

\begin{figure*}{}
\resizebox{2.0\columnwidth}{!} {\includegraphics{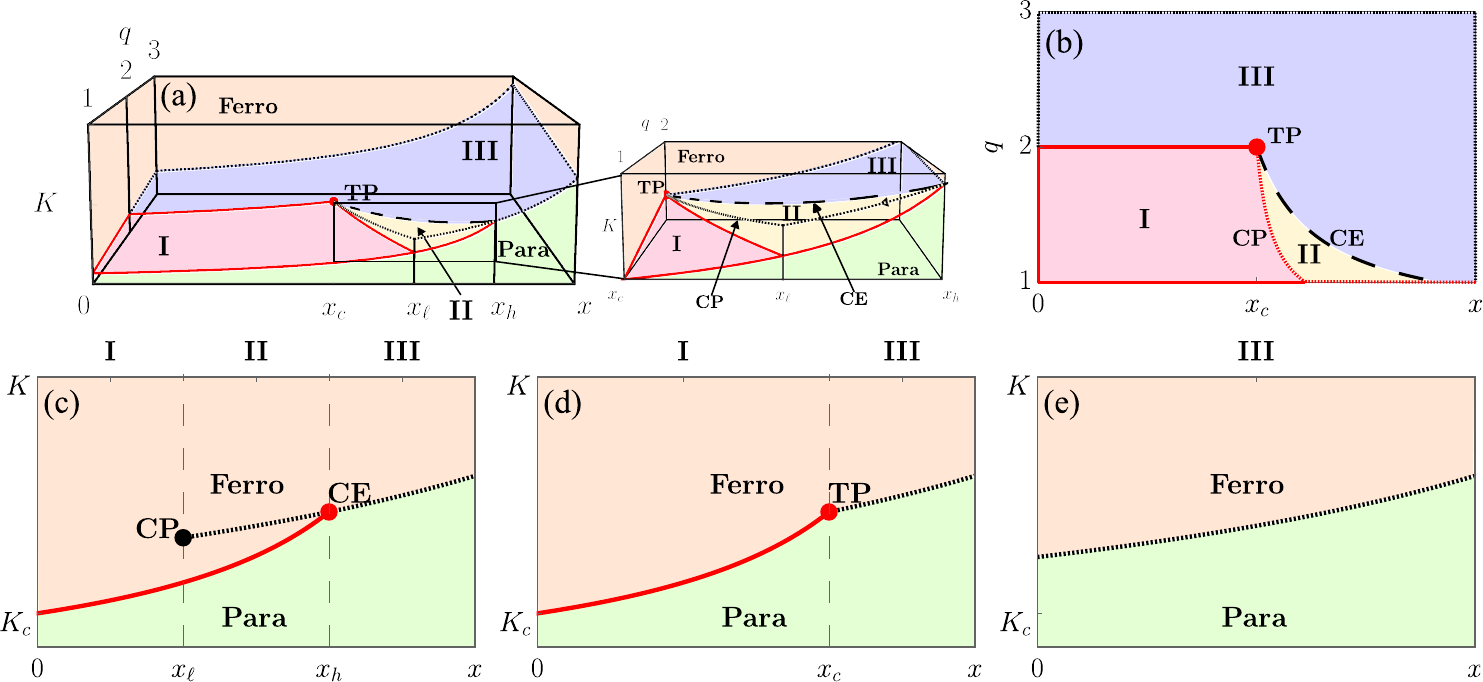}}
\caption{ 
Phase schematic of QR-Potts model. 
(a) Three-dimensional phase diagram in parameter space ($x=r/q$, $q$, $K = T^{-1}$). I represent a region with $x\in [0,x_\ell]$, where the second-order transition occurs at $T_c$ from para to ferro phases for fixed $x$ and $q$. II represents a region with $x\in [x_\ell, x_h]$, where second-order transition $T_c$ and first-order transition at $T_f$ occur subsequently as $T$ decreases. III represents region with $x\in [x_h, \infty]$, where first-order transition occurs at $T_f$. (b) Phase diagram in space ($x$,$q$). They display phase transition type in parameter space $[q, r]$. Solid and dotted lines and dashed lines represent continuous and discontinuous transitions. CP and CE represent critical points and endpoints, which can be illustrated more clearly in (c). 
TP is located at $q=2$ and $x_c=2e/3$ at which CP and CE lines are merged. CP line reaches $x_{1\ell}=e^2$ as $q\to 1$.
CE line reaches $x_{1h}\simeq 9.60$ as $q\to 1$. 
(c) Phase diagram in space $[x,K=T^{-1}]$ for $1 \le q<2$. Various types of phase transitions occur depending on $x=r/q$.
(d) When $q=2$ and $x\le x_c$, continuous transition occurs, whereas when $x > x_{c}$, discontinuous transition occurs.
(e) When $q>2$, first-order transition occurs regardless of $x$ value.
\label{fig:fig1}	
} 
\end{figure*}

Here, we employ the Ginzburg–Landau (GL) formalism to the QR-Potts model to reveal a much richer phase diagram, as illustrated in Fig~\ref{fig:fig1}, than the one known before~\cite{tamura_2010, Okumura, tanaka_2011, Krasnytska}. The two lines comprising critical points (CP) and the critical endpoints (CE), which merge at a tri-critical point (TP), divide the parameter space into the regimes exhibiting continuous, discontinuous, consecutive, and hybrid transitions, respectively, in the parameter space. The GL formalism allows us to trace the stable and metastable states, revealing physical mechanisms underlying the consensus formation and the influence of shy voters. These analytic results provide a deeper understanding of the critical phenomena than numerical results. Interestingly, we find that the phase diagram of the hidden Potts model is similar to that of the AT model on scale-free networks~\cite{AT, MS}, in which two types of Ising spins located on each node are subject to intra- and inter-type interactions. This similarity suggests that the results obtained here can be universal to various systems. 

The paper is organized as follows: In Sec.~\ref{sec:model and formalism}, we introduce the GL free energy of the QR-Potts model and present the method to obtain two order parameters $m$ and $m_r$, representing the fraction of spins in a given visible state and {\it any} hidden states, respectively. In Sec.~\ref{sec:OVS}, we employ the analytic and numerical methods based on Sec.~\ref{sec:model and formalism} to obtain the phase diagram illustrating different phase transitions of the order parameter $m$. The other order parameter for hidden states is studied in Sec.~\ref{sec:OHS}. The implication of our results for social opinion dynamics is discussed in Sec.~\ref{sec:opinion}. The results are summarized in Sec.~\ref{sec:summary}.

\section{Model and Formalism}
\label{sec:model and formalism}

Approximating the two-spin interaction in ~\ref{eq:hamiltonian_Hidden} in terms of interaction with MFs, valid for spatial dimension larger than upper critical dimension, one can obtain the GL free energy, which allows further analysis, numerical and analytical, for the QR-Potts model. We introduce the average probability of a spin to be in the first visible state, the other visible states, and hidden states as
\begin{align}
\langle \delta_{s_i, \alpha} \rangle =
\left\{ 
\begin{array}{lll}
\mu &~~\textrm{for}~ &~~\alpha = 1, \cr
\nu_{1} &~~\textrm{for}~ &~~\alpha = 2, \cdots, q, \cr
\nu_{2} &~~\textrm{for}~ &~~\alpha = q+1, \cdots, q+r,
\end{array} 
\right.
\label{eq:mf}
\end{align}
respectively, which will play the role of the MFs. Notice that $\mu+(q-1)\nu_1$ and $r \nu_2$ are the probability that a spin is in visible and hidden states. They satisfy $ \mu+ (q-1) \nu_{1} + r \nu_{2} = 1$, allowing us to represent $\nu_2$ in terms of $\mu$ and $\nu_1$. 

The partition function, $\mathcal{Z} \equiv \sum_{\{s_i\}} e^{-\beta \mathcal{H}}$, where $\beta\equiv 1/k_BT$ with the Boltzmann constant $k_B$, can be decomposed as $\mathcal{Z}\simeq \sum_{\mu, \nu_1} \mathcal{Z}_{\mu, \nu_1}$ with the constrained partition function $\mathcal{Z}_{\mu, \nu_1} \equiv \sum_{\{s_i\}}^\prime e^{-\beta \mathcal{H}}$, where the prime indicates that the summation runs under the constraint. Neglecting the quadratic fluctuations such as $(\delta_{s_i,\alpha} - \langle \delta_{s_i,\alpha}\rangle)(\delta_{s_j,\alpha} - \langle \delta_{s_j,\alpha}\rangle)$, one obtains the effective MF Hamiltonian 
\begin{align}
& -\beta\mathcal{H}^{\textrm{mf}}_{\mu, \nu_{1}} = 
    - K\sum_{\langle i,j \rangle}\left(\mu^2+\sum_{\alpha = 2}^{q}\nu_1^2\right) \cr
& \hspace{20pt}
 +K\sum_{i} \sum_{j\in \text{n.n.} (i)} \left( (h + \mu) \delta_{s_i, 1} + \sum_{\alpha = 2}^{q} \nu_{1} \delta_{s_i, \alpha} \right),
\label{eq:hamiltonian_Hidden_meanfield}
\end{align}
where $K\equiv \beta J = 1/T$ (we set $J / K_B \equiv 1$), $\text{n.n.}(i)$ is the set of the nearest neighbors of site $i$, and $h\equiv H/(Jz)$ is the re-scaled external field. We assume that every site has the same number $z$ of the nearest neighbors. The free energy $f_{\mu,\nu_1}^{\rm mf} 
\equiv -N^{-1} \ln \mathcal{Z}_{\mu,\nu_1}^{\textrm{mf}}$ is obtained from the partition function $\mathcal{Z}_{\mu,\nu_1}^{\textrm{mf}} = \sum_{\{s_i\}} e^{-\beta\mathcal{H}^{\textrm{mf}}_{\mu, \nu_{1}}}$ in the MF limit.

Our main interest covers two issues: (i) whether spins are more likely to align in the direction of the first state, and (ii) how likely a spin is in hidden states as a function of inverse temperature $K$. 
Therefore we consider
\begin{align}
m\equiv \mu-\nu_{1} \text{ and } m_r\equiv r\nu_{2}
\end{align}
instead of $\mu$ and $\nu_1$. $m$ is close to $1$ if a spin is most likely to be in the first state and $0$ if it is equally likely in all the visible states or most likely in hidden states. $m_r$ distinguishes the latter two cases, representing the probability of being in hidden states. Rewriting $f_{\mu,\nu_1}^{\rm mf}$ as a function of $m$ and $m_r$, we obtain the GL free energy 
\begin{align}
f(m,m_r,h) \simeq \dfrac{1}{2\tau}\left\{ (1-m_r)^2 + (q-1) m^2\right\} - \ln \mathcal{X},
\label{eq:fmR}
\end{align}
where,
\begin{align}
\mathcal{X} = e^{(h + 1 - m_r +(q-1)m)/\tau} + (q-1)e^{(1 - m_r - m)/\tau} + r.
\end{align}
Here, we introduce a re-scaled temperature $\tau\equiv {q/K z} = {q T/ z}$ and call it temperature hereafter. We set $h\to 0^+$ unless stated otherwise and denote the GL free energy simply by $f(m,m_r)$.

In the thermodynamic limit $N\to\infty$, the GL free energy of the whole system in the MF limit can be approximated by $\bar{f} \equiv -N^{-1}\ln \mathcal{Z} \simeq \min_{m,m_r} f(m,m_r)$ and the global minimum location gives the order parameters $(\bar{m},\bar{m}_r)=\arg\min_{m,m_r} f(m,m_r)$. They satisfy ${\partial f}/{\partial m}\big|_{\bar{m},\bar{m}_r}={\partial f}/{\partial m_r}\big|_{\bar{m},\bar{m}_r}=0$, yielding 
\begin{align}
\bar{m}= {{e^{q\bar{m}/\tau} - 1} \over {e^{q\bar{m}/\tau} + q-1 + r e^{-(1-\bar{m}-\bar{m}_r)/\tau}}},
\label{eq:equationOfState_m_Hidden}
\end{align}
and
\begin{align}
\bar{m}_r={r e^{-(1-\bar{m}- \bar{m}_r)/\tau}\over e^{q\bar{m}/\tau} + q-1 + r e^{-(1-\bar{m}- \bar{m}_r)/\tau}}.
\label{eq:equationOfState_R_Hidden}
\end{align}

Practically, we first obtain $m_r(m)$ from ~\ref{eq:equationOfState_R_Hidden} and utilize it in ~\ref{eq:fmR} to define a single-variable function $f(m) \equiv f(m,m_r(m))$. Then, the order parameters are given by
\begin{equation}
\bar{m} = \arg\min_m f(m) \ {\rm and} \ \bar{m}_r = m_r(\bar{m}).
\label{eq:argmin}
\end{equation}
Also, the expansion of $f(m)$ near $m=0$ can offer an analytic understanding of the order parameters. We first expand ~\ref{eq:equationOfState_R_Hidden} with respect to $m$ to obtain $m_r(m)=m_r(0)+\sum_{n=2}^\infty B_n \left(\dfrac{q m}{\tau}\right)^n$ with 
$B_n$'s constants and inserting it into ~\ref{eq:fmR}, one finds
\begin{align}
f(m) =\sum_{n=0}^\infty C_n \left(qm \over \tau \right)^n.
\label{eq:free_energy_density}
\end{align}
The coefficients $B_n$'s and $C_n$'s are in Supporting information (SI.~\ref{seca:Cn}. Hereafter, we drop the bar notation in $\bar m$ and $\bar m_r$ for simplicity.

\begin{figure*}
\resizebox{1.0\linewidth}{!}{\includegraphics{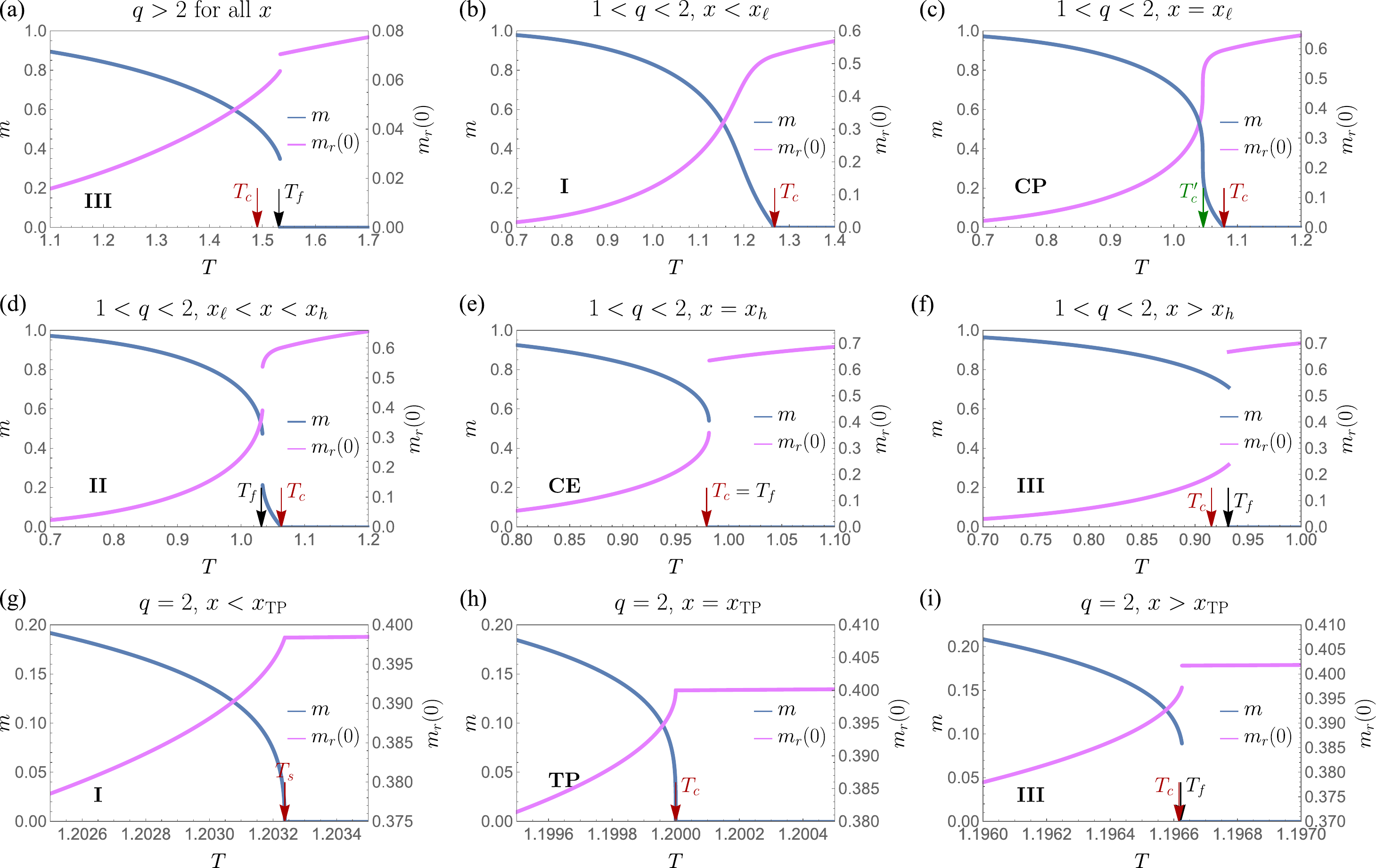}}
\caption{
Plot of order parameters $m$ (left axis) and $m_r$ (right axis) vs temperature $T = z \tau / q$ where $z=4$ for all the cases: (a)--(e) for $1 < q = 1.5 < 2$, and (a) $x<x_\ell$, (b) $x = x_\ell$, (c) $x_\ell < x < x_h$, (d) $x = x_h$, and (e) $x > x_h$. (f)--(h) at $q=2$ and for (f) $x<x_c$, (g) $x = x_c$, and (h) $x > x_c$. (i) For $q = 2.5 > 2$ and $x>0$. Continuous transitions occur in (a)–(c), (f), and (g). Discontinuous transitions occur in (c)–(e), (h), and (i). A hybrid transition occurs in (d). Note that in (b), another type of continuous transition occurs at $T_c^{\prime}$ in which $\partial^2 f(m, T) / \partial m^2 \to 0$. This implies $\partial m / \partial T \to \infty$.}
\label{fig:fig2}	
\end{figure*}

\begin{figure}
\centering
\resizebox{0.8\columnwidth}{!}{\includegraphics{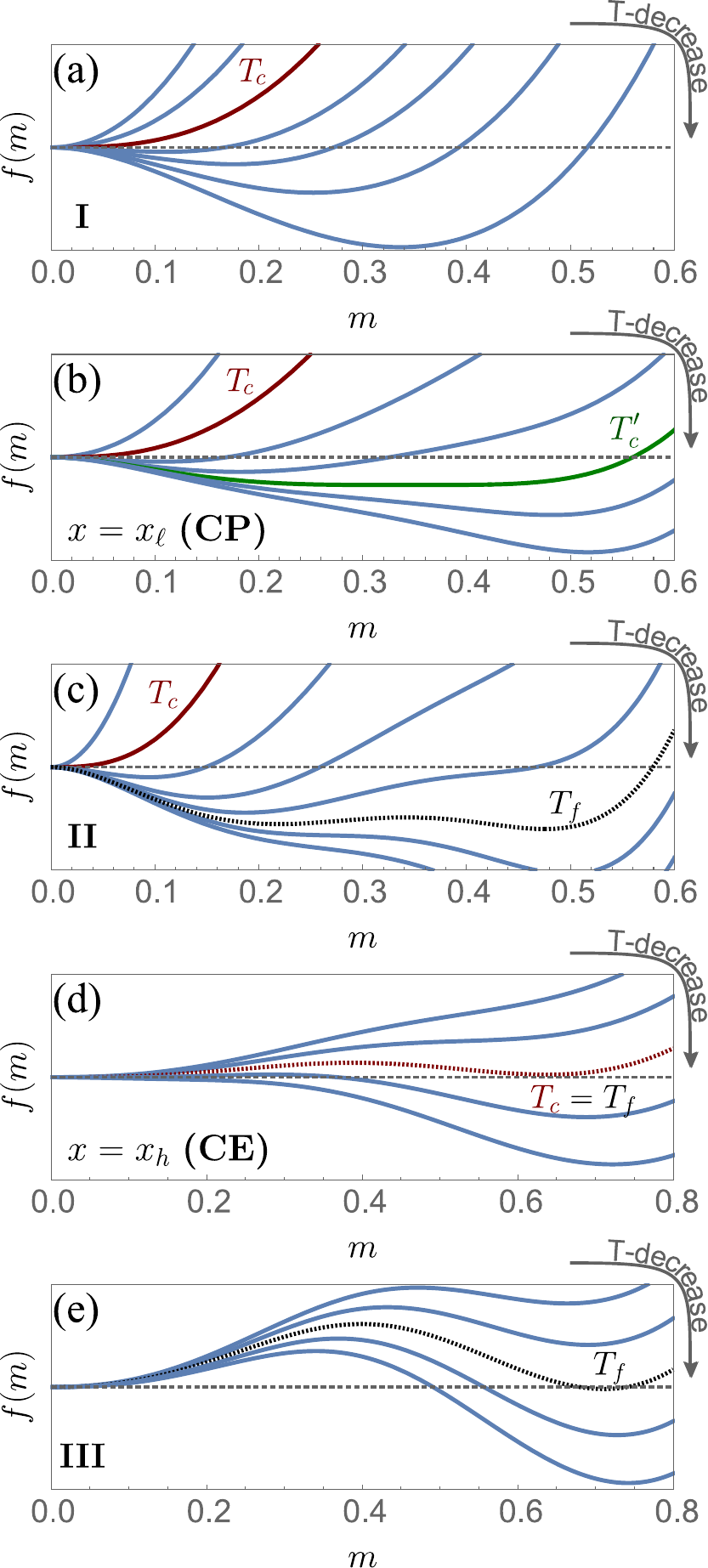}}
\caption{ Plots of free energy densities $f(m)$ vs $m$ for several cases of $1< q=1.5 < 2$, where $f(m)$ is reset to zero when $m=0$ for illustration. (a) A continuous transition occurs in Regime I with $x=3.0 < x_\ell$. At $T_c$, $\partial^2 f(m)/\partial m^2=0$ at $m=0$. As $T$ decreases below $T_c$, the minimum of $f(m)$ decreases continuously. Thus, a continuous transition occurs. 
(b) At $x=x_\ell\approx 4.0$, global minimum of $f(m)$ occurs near $m=0$ at $T_c$ for second-order transition. As $T$ decreases and reaches $T_c^\prime$, position $m$ of global minimum increases rapidly but continuously. Order parameter increases rapidly, as presented in Fig.~\ref{fig:fig2}(b). 
(c) When $x_\ell=4.1 < x < x_h$, global minimum of $f(m)$ occurs at $m=0$ at $T_c$ for second-order transition. As $T$ decreases below $T_c$, minimum position $m_1$ slowly increases and becomes finite value $m_1$. Order parameter increases continuously. As $T$ decreases further, another minimum occurs at $m_2$. At $T_f$, $f(m_1)=f(m_2)$ and discontinuous transition occurs at $m_2$. Order parameter jumps from $m_1$ to $m_2$.
(d) At $x=x_h\approx 4.685$, $T_c=T_f$. The first and second derivatives of $f(m)$ at $m=m_1=0$ are zero. So, susceptibility diverges at $T_c$. At the same temperature, another minimum occurs at $m_2$ as $f(m_1)=f(m_2)$. Thus, the discontinuous transition occurs. Hybrid transition occurs. This point is referred to as CE. 
(e) For $x = 5.20 > x_h$, discontinuous transition occurs at $T_f$. 
\label{fig:fig3}	
} 
\end{figure}

\section{Phase Transition for visible states}
\label{sec:OVS}

In this section, we utilize both the numerical solutions [~\ref{eq:argmin}] and the analytic expansion [~\ref{eq:free_energy_density}] with the coefficients given in Supplementary Information (SI)~\ref{seca:Cn} to trace the global minimum of $f(m)$ varying with $\tau, q,$ and $x$ to obtain the phase diagram [Fig.~\ref{fig:fig1}]. The change of the sign of $C_3$ at $q=2$ [SI.~\ref{seca:Cn}] is responsible for different types of phase transitions between $q<2$ and $q>2$ in the Q-Potts model ($r=0$)~\cite{Wu} and also in the QR-Potts model. Therefore, we consider the cases of $q > 2$, $1 < q < 2$, and $q=2$. 

\subsection{$q>2$: Discontinuous transition}

When $q$ is close to and larger than $2$, the fourth-order expansion $f(m)$ in ~\ref{eq:free_energy_density} with $C_3\propto -(q-2)$ and $C_2, C_4>0$ can approximate well $f(m)$ and reveal the change of the global minimum at temperature $\tau_f$ [Fig.~\ref{fig:fig2}(a)], which is given for small $x$ by $\tau_f \simeq \tau_{f}^{(0)} - {x\over e}$ with $\tau_f^{(0)} \simeq 1 + {(q-2)^2 \over 6(q-1)}$ and the order parameter
\begin{equation}
{m} \simeq 
\begin{cases}
0 
& \ {\rm for} \ \tau>\tau_f,\\
\tau_{f}\dfrac{q-2}{q-1} \left(1 + {3x\over 2e} \right)
& \ {\rm for } \ \tau<\tau_f.
\end{cases}
\label{eq:DPTsmallr}
\end{equation}
The transition temperature $\tau_f$ decreases as $x$ increases because the energetic effect, favoring the ordered state, can dominate the entropic effect, favoring hidden states only at low temperatures as $x$ increases.

\subsection{$1<q<2$: Continuous, successive, hybrid, and discontinuous transitions}
When $1 < q < 2$, $C_3>0$ but $C_4$ varies with $x$, which underlies different types of phase transitions occurring in three regimes, denoted as I for $x < x_\ell(q) $, II for $x_\ell(q) < x < x_h(q)$, and III for $x > x_h(q)$ with the CP $x_\ell(q)$ and at the CE $x_h(q)$ [Fig.~\ref{fig:fig1}(c)]. 

\noindent {\it $-$ Continuous transition:} 
In Regimes I and II ($x<x_h$), the global minimum of $f(m)$ at $m_1$ is shifted continuously from zero to an infinitesimally positive value as temperature decreases passing the critical temperature $\tau_c$ [Figs.~\ref{fig:fig2}(b) and (d), and ~\ref{fig:fig3}(a) and (c)]. It is related to the change of the sign of $C_2$ at $\tau_c$, leading to $\tau_c = {1\over 1+xe^{-1}}$ for $x<e^2$. Near $\tau_c$, $m$ is zero or quite small, and one can employ the third-order expansion in ~\ref{eq:free_energy_density} with $C_2\propto \tau-\tau_c$ to find that 
\begin{equation}
{m} \simeq \begin{cases}
0 & \ {\rm for} \quad \tau > \tau_c,\\
\dfrac{2}{2-q}\dfrac {\tau_c-\tau}{\tau_c} & \ {\rm for} \quad
\tau<\tau_c.
\end{cases}
\label{eq:CPTsmallr}
\end{equation}

\noindent {\it $-$ Discontinuous transition:} 
In Regime II and III ($x>x_\ell$), $C_3$ is positive but $C_4$ is negative such that the free energy $f(m)$ increases with $m$ owing to $C_3>0$ and then decreases owing to $C_4<0$ and increases again because of positive higher-order terms, resulting in two local minima at $m_1$ and $m_2$ and the change of the global minimum between them at a temperature $\tau_f$ [Fig.~\ref{fig:fig3}(c)-(e)]. The discontinuous transitions in these regimes are presented in Fig.~\ref{fig:fig2}(d) (II) and Fig.~\ref{fig:fig2}(f) (III). 

Because a continuous transition also occurs in Regime II, one finds double transitions in Regime II [Fig.~\ref{fig:fig1}(c)]: As the temperature decreases, the order parameter $m$ changes from zero to an infinitesimally positive value at $\tau_c$ and then discontinuously jumps to a finite positive value ($m_2$) at $\tau_f$, which is less than $\tau_c$. 

In Regime II, $\tau_c$ is higher than $\tau_f$. As $x$ increases, however, $\tau_c$ decreases faster than $\tau_f$, and therefore, they meet at $x_h$, called the CE. For $x>x_h$ (Regime III), the continuous increment of the local minimum $m_1$ at the temperature $\tau_c$ less than $\tau_f$ does not affect the global minimum maintained at $m_2$. Therefore, one can see just a discontinuous transition at $\tau_f$. 
 
\noindent {\it $-$ Explosive hybrid transition at CP ($x_\ell$):}
At $x=x_\ell$ (CP), no discontinuous jump is seen [Fig.~\ref{fig:fig2}(c)], but as the temperature is further lowered than $\tau_c$, the order parameter $m$ increases continuously but explosively [Fig.~\ref{fig:fig2}(c)] near the characteristic temperature $\tau_c^{\prime}$, i.e, $dm/d\tau|_{\tau_c^{\prime}}\to\infty$. 
As temperature is near $\tau_{c}^{\prime}$, $m$ increases continuously as
\begin{align}
 m-m_{c}^{\prime}\sim (\tau_c^\prime-\tau)^{\beta^\prime} 
 \label{eq:CP_Criticality}
\end{align} with $\beta^\prime=1/2$. The detailed derivation is presented in SI.~\ref{seca:CP_Crit}. Hence, a hybrid transition occurs. At $m_c^\prime$, $\partial f/\partial m=\partial^2 f/\partial^2 m=0$ as shown in Fig.~\ref{fig:fig3}(b). Therefore, critical behavior appears at $(\tau_c^\prime, m_c^\prime)$. 

\noindent {\it $-$ Hybrid transition at CE ($x_h$):} 
At $x=x_h$ (CE), the change of the global minimum from $m_1=0$ to $m_2>0$ and the gradual shift of the former local minimum from $m_1=0$ to finite $m_1$ coincide at $\tau_f=\tau_c$ [Figs.~\ref{fig:fig2}(e) and ~\ref{fig:fig3}(d)]. Interestingly, the susceptibility $\chi \equiv {\partial m\over \partial h}$ is finite in the side $\tau\to \tau_c^-$ but diverges in the other side  $\tau\to \tau_c^+$ in Fig.~\ref{fig:fig5} in SI.~\ref{seca:chi}. It implies that the transition is hybrid or mixed-order, exhibiting the properties of continuous and discontinuous transitions simultaneously~\cite{HPT1, HPT2, HPT3, HPT4}.
We reveal that $\partial f/\partial m=0$ at $m=m_1=0$ and $m=m_2 > 0$ and $\partial^2 f/\partial^2 m=0$ only at $m=m_1=0$ as shown in Fig.~\ref{fig:fig3}(d). Therefore, critical behavior appears at $(\tau_c^+,0)$. 

\subsection{$q=2$: Continuous and discontinuous transitions}
\label{subsec:q2}
When $q=2$, the CP and CE lines are merged at $x_{\rm TP}$. See Fig.~\ref{fig:fig1}(b) and SI.~\ref{seca:TP}. The free energy is expanded as $f(m) \simeq C_2 (qm/\tau)^2 + C_4 (qm/\tau)^4 + C_6 (qm/\tau)^6$.
When $x < x_{\rm TP}$, $C_4 > 0$, and a single minimum at $m_1=0$ is shifted continuously to finite $m_1$ at temperature near $\tau_c$ [Fig.~\ref{fig:fig2}(g)]. Because $C_2 \sim \tau-\tau_c$, one finds $m\sim (\tau_c-\tau)^{1/2}$ for $\tau<\tau_c$, with the critical exponent $\beta=1/2$ different from $\beta=1$ in ~\ref{eq:CPTsmallr} for $1<q<2$. When $x > x_{\rm TP}$, $C_2 > 0$ but $C_4 <0$, and thus $f(m)$ can have multiple minima over a wide range of temperatures. A discontinuous transition occurs at $\tau_f$ [Fig.~\ref{fig:fig2}(i)], which is approximately $\tau_c + {C_4^2\over 4 C_6}$ when the finite jump $\Delta m\simeq \sqrt{|C_4|\over C_6}$ at $\tau_f$ is small. Therefore, one finds that $\tau_c=\tau_f$ and $\Delta m=0$ if $C_4=0$, which holds at $x=x_{\rm TP}={2e/ 3}$ and $\tau = \tau_{\rm TP}={3/5}$ [SI.~\ref{seca:TP}], implying that it is a TP. At $x_{\rm TP}$, the free energy is given by $f(m)\simeq C_2(qm/\tau)^2 + C_6 (qm/\tau)^6$, and therefore $m$ indicates a continuous transition at $\tau_c$ [Fig.~\ref{fig:fig2}(h)] as
\begin{equation}
{m} \sim \begin{cases}
0 & \ {\rm for} \quad \tau > \tau_{\rm TP},\\
\left(\tau_{\rm TP} -\tau\right)^{1/4} & \ {\rm for} \quad \tau<\tau_{\rm TP}.
\end{cases}
\label{eq:CPTatTP}
\end{equation}

Near TP, the CP and CE lines exist, exhibiting a super-criticality: $x_{\ell}-x_{\rm TP} \sim (2-q)^{2/3}$ along the CP line and $x_{h}-x_{\rm TP} \sim (2-q)^{2/3}$ along the CE line. The sixth-order expansion of $f(m)$ is adopted to derive these results as presented in SI.~\ref{seca:TP}. Therefore, we find that the gap $\Delta x \equiv x_h-x_\ell$ shrinks as $\Delta x \sim (2-q)^{2/3}$. Note that this gap is not the discontinuity of the order parameter as revealed in the conventional critical behavior, such as the liquid–gas transition, but the discontinuity between the two transition lines $x_\ell$ and $x_h$. Moreover, this novel supercritical behavior appears in the AT model with the same critical exponent value. Interestingly, the CP and CE lines in the AT model with $q$ states (Q-AT) on the random graph also exhibit the supercritical behavior with the exponent $2/3$, suggesting a super-universality of the QR-potts and Q-AT model in the MF limit. 

The phase diagram of the QR-Potts model spanned by $(q, x=r/q, 1/T)$ is similar to that of the AT model on scale-free (SF) networks~\cite{AT, MS}. The phase diagrams in Fig.~\ref{fig:fig1}(b) and (c) appear similar to Fig. 2(a) in Ref.\cite{AT}, commonly displaying the CE and CP lines. The parameter space of the AT model on SF networks is spanned by $(\lambda, x_{\rm AT}, 1/T)$. $\lambda$ is the degree exponent of the SF network and $x_{\rm AT}$ is the intra- and inter-type interaction ratio. From the perspective of the phase diagram, $(\lambda,x_{\rm AT})$ corresponds to ($q, x=r/q$) in the QR-Potts model. The correspondence between ($\lambda, x_{\rm AT}$) and ($q,x$) seems non-trivial. However, it is clear that quantities ($\lambda, x_{\rm AT}$) affect both the interaction energy and entropy to change the free energy landscape as ($q,x$) in the QR-Potts model.

\section{Phase Transition for the hidden states in $q\to 1^+$}
\label{sec:OHS}

\begin{figure*}
\resizebox{2.0\columnwidth}{!}{\includegraphics{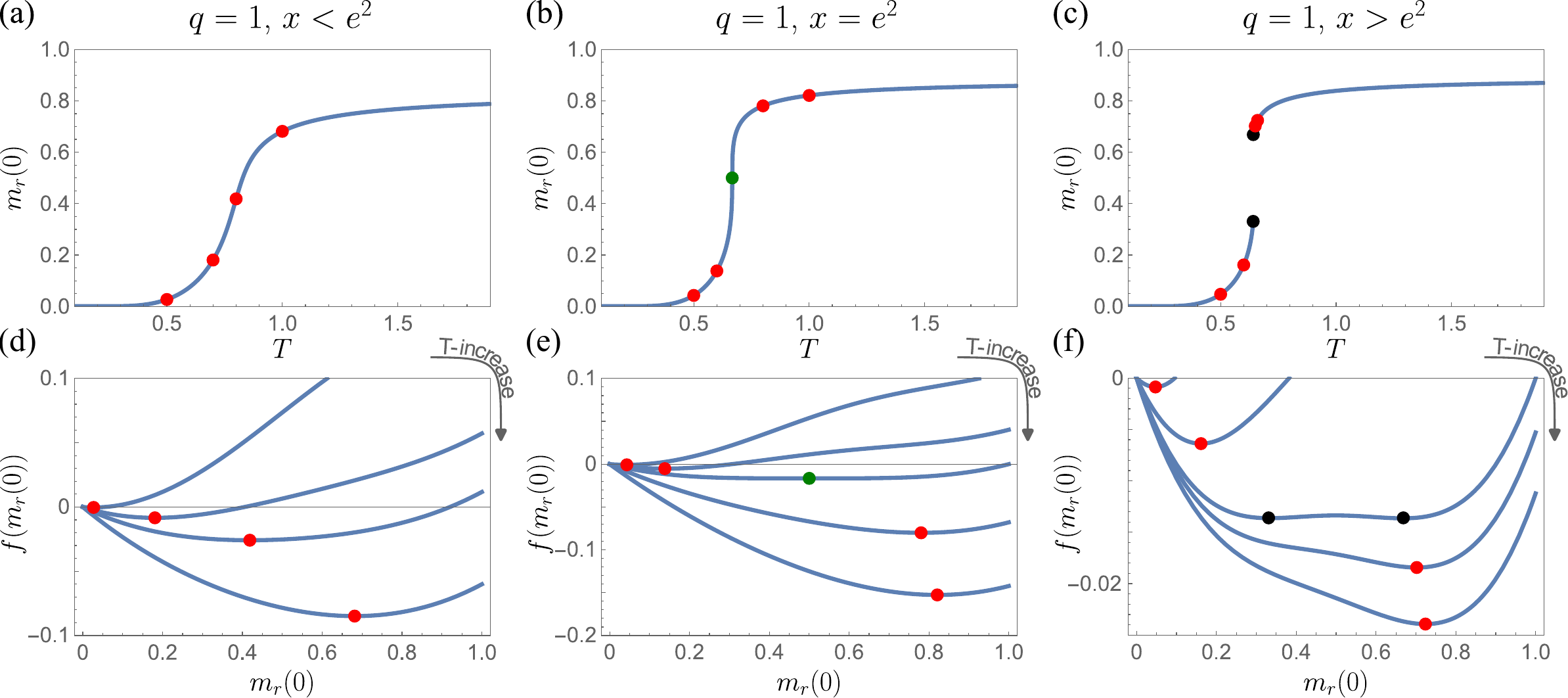}}
\caption{
(a)–(c) Plots of $m_r(0) \equiv y$ vs $T$ for various values of $x$ but with fixed $q=1$ $\in(1,2)$:(a) when $x<e^2$, (b) when $x=e^2$, and when $x>e^2$. 
(d)–(f) Plots of {\it ad hoc} potential vs $y$ for various values of $x$: (d) when $x<e^2$, (e) $x=e^2$, and (f) $x>e^2$. The green (dark) circle point in (b) is located at $\tau_c^{\prime}$, at which $d^2f(m_{r})/dm_{r}^2 |_{m_{r}=m_{r}^{*}}=0$ ($\tau_c^{\prime} = m_{r}^{*}(1-m_{r}^{*})$). Hence, the potential is flat at $m_{r}^*(\tau_c^{\prime})$. This point corresponds to the TP of a discontinuous percolation in the limit $q\to 1$.}
\label{fig:fig4}	
\end{figure*}

The Kasteleyn-Fortuin mapping~\cite{KF_Percolation_1,KF_Percolation_2} relates the Q-Potts model to the bond percolation in the limit $q \to 1$. It is still valid for the QR-Potts model. 
In the QR-Potts model with $q\to 1^+$, a discontinuous transition of $m$ occurs for $x > x_{1\ell}=e^2$ [Fig.~\ref{fig:fig1}(c)]. With such large $x$, there can be multiple solutions $m_r$ of ~\ref{eq:equationOfState_R_Hidden}, resulting in a discontinuous jump of $m_r$, which can drive the discontinuous transition of $m$ for $q\to 1^+$. 

Consider the GL free energy in ~\ref{eq:fmR} with $q\to1$, 
\begin{align}
f(m,m_{r}) \to f(m_{r}) = {m_{r}^2-1\over 2\tau} - \ln (q + r e^{-{1-m_{r}\over \tau}}).
\end{align}
Its global minimum location $\bar{m}_{r} = \arg \min_{m_{r}} f(m_{r})$ satisfies $\partial f(m_{r}) / \partial m_r|_{m_{r}=\bar{m}_{r}} \equiv f_{m_r}(m_{r})|_{m_{r}=\bar{m}_{r}}=0$ or equivalently
\begin{equation}
\bar{m}_{r}={xe^{-(1-\bar{m}_{r})/\tau}\over 1+ x e^{-{(1-\bar{m}_{r})/\tau}}}.
\label{eq:F}
\end{equation}
If $x<x_{1\ell}=e^2$, $f(m_{r})$ has a single minimum at $\bar{m}_{r}$ for all $\tau$ [Fig.~\ref{fig:fig4}(a)]. If $x > x_{1\ell}$, two local minima $\bar{m}_{r-}$ and $\bar{m}_{r+}$ exist for $\tau_-<\tau<\tau_+$ [Fig.~\ref{fig:fig4}(c)]. Notice that at $x_{1\ell}=e^2$, $f_{m_r} = f_{m_r m_r}=f_{m_r m_r m_r}=0$ at $m_{r}={1\over 2}$ and $\tau= {1\over 4}$ [Fig.~\ref{fig:fig4}(b)]. The range of temperature having both $\bar{m}_{r-}$ and $\bar{m}_{r+}$ can be obtained by using the condition $f_{m_r} = f_{m_r m_r}=0$ at $m_{r}=\bar{m}_{r\pm}$ and $\tau=\tau_\mp$ yielding $\tau_\mp=\bar{m}_{r\pm} (1-\bar{m}_{r\pm})$ and ~\ref{eq:F} for given $x$. When $x$ is large, $\bar{m}_{r-}\sim 1/\ln (x\ln x)$ and $\bar{m}_{r+}\sim 1- e/x$ and therefore is $m_r$ behaves as
\begin{equation}
m_r \simeq 
\begin{cases} 
{x} e^{-{1}\over{\tau}} & \ {\rm for} \ \tau \ll \tau_{f}^{\prime},\\
{1 \over \ln \left( {x} \ln x \right)}& \ {\rm for} \ \tau \to \tau_{f}^{\prime-},\\
1 - {e\over x} & \ {\rm for} \ \tau > \tau_{f}^{\prime},
\end{cases}
\label{eq:mr0}
\end{equation} 
where $\tau_{f}^\prime$ is the temperature located between $\tau_-$ and $\tau_+$ [SI.~\ref{seca:OHS}].

When $q\to 1^+$, ~\ref{eq:equationOfState_R_Hidden} is independent of $m$, indicating that the solution of ~\ref{eq:F} can be utilized in place of $m_r$ in ~\ref{eq:equationOfState_m_Hidden} to determine $m$ as 
\begin{align}
m=1-e^{-m/\tau}-mxe^{-(1-m_{r})/\tau}.
\label{eq:BP}
\end{align}
If $r=0$, then ~\ref{eq:BP} reduces to $m =1-e^{-m/\tau}$, known for the bond percolation in random graphs~\cite{Newman}. 
From ~\ref{eq:BP}, one can obtain the features of bond percolation transition for the QR-Potts model. Consequently, in case $m_{r}$ is discontinuous, $m(r\ne 0)$ exhibits a discontinuous transition. 

\section{Discussions: Potential applications to social opinion formation}
\label{sec:opinion}

The QR-Potts model renders intriguing insights into the interplay between conformity and individualism in social dynamics and opinion formation. The QR-Potts model explores how individuals in a social network choose between following the majority opinion and keeping their ideas hidden. The model reveals that this competition can lead to a sudden emergence of consensus or the prevalence of hidden opinions under certain conditions. Suppose every agent has two options: following the majority opinion or selecting a random opinion. The former option contributes to the free energy as much as $qe^{\beta J}$, whereas the latter option incurs a free energy cost of $q+r$. Thus, there exists a characteristic temperature $T_\circ \approx 1/\ln (1+ r/q)$ such that when the strength of social fluctuation $T$ is lower (higher) than $T_\circ$, each agent should take the ordered (hidden) opinion. If the number of hidden states $r$ is much larger than unity, then the characteristic temperature $T_\circ$ is much lower than $1$. This implies that even at relatively low temperatures, agents are prone to hide their opinions rather than follow to the majority. 

When the system contains many hidden states ($x > e^2$), the order parameter $m_r(m)$ exhibits bi-stable states [SI.~\ref{seca:V_HS}], including a hidden super-cooling state between them. As we know, in the liquid–gas transition, the system remains in a meta-stable state for an extended period. This implies that an unstable, highly disordered social system, akin to a hidden super-cooling state, can endure significant social suppression. This resilience is particularly striking given high suppression factors, such as authoritarian regimes (in the limit $q\to 1^+$) or oppressive political conditions.

Comparing swing agents with zealots~\cite{Mobilia1, Mobilia2, Mobilia3, Mellor, Khalil1, Khalil2} in the context of opinion dynamics provides valuable insights into how two different types of agents affect the dynamics of opinion propagation. swing agents and zealots can be mirror images of each other, with one group actively influencing others and the other group remaining passive and unresponsive. Developing an opinion dynamics model that accommodates both swing agents and zealots can offer a better understanding of how diverse agent behaviors shape the evolution of opinions within a society.

Finally, we note that shy voters change their states because of thermal fluctuations as noise-induced updating in the noisy voter model~\cite{Khalil1, Khalil2, Carro, Peralta, ARtime}, where an opinion changes spontaneously and independently of the neighbors' state. 
However, the shy voter in our model affects the entropy change of the system, followed by the change of the free energy landscape. Therefore, the state updating in the shy voters cannot be regarded as a finite-size effect. In the continuum limit, the shy voters do not behave randomly; however, they make a new local minimum of the system in the thermodynamic limit. Recently, a variant of the voter model has been shown to display a change in entropy production~\cite{Oliveira_Entropy_Production}, leading to a change of entropy and free energy landscape.

\section{Summary}\label{sec:summary}

We studied the phase transition of the QR-Potts model comprising $q$ visible states and $r$ hidden states. The Hamiltonian includes the interactions between two spins in visible states but no interaction with a spin in hidden states. Applying the Ginzburg–Landau formalism in the MF limit, we obtained a rich phase diagram including a variety of phase transitions, such as continuous, discontinuous, and hybrid (or mixed-order) transitions, and diverse kinds of transition points such as CP, CE, and TP in the parameter space comprising temperature $T$, the number of visible states $q$, and the ratio $x=r/q$. This phase diagram is close to that of the AT model, implying that the two types of spin states in the QR-Potts model play the role of two layers of different spin species in the AT model.

The phase transition of the QR-Potts model highly depends on $q$ and $x$. We analytically uncovered diverse features of the phase transitions. When $q > 2$, a discontinuous transition occurs regardless of the value of $x$. When $1 < q < 2$, the transition is continuous in Regime I ($x < x_\ell$) while it is discontinuous in Regime III ($x > x_h$). Between the two regimes, ($x_\ell < x < x_h$) called Regime II, a continuous and then a discontinuous transition occurs consecutively as $T$ decreases. Along the curve, $x_\ell(q)$ for $1 \le q < 2$, the order parameter increases continuously with decreasing $T$ lower than $T_c$, but its derivative with respect to temperature diverges at $T_c^\prime$. As temperature is near $T_c^\prime$, the order parameter behaves $m-m_c\prime \sim (T_c^\prime-T)^{\beta^\prime}$. So, the singular behavior of the hybrid transition occurs near $m_c > 0$. On the other hand, along the CE curve at $x_h(q)$ for $1 \le q < 2$, another type of hybrid transition occurs, in which the order parameter is discontinuous, but the susceptibility $\partial m/\partial h$ diverges at $m=0$. Note that the hybrid transition at the CP line can be found in other non-equilibrium models such as $k$-core percolation~\cite{kCore_Lee} and the hybrid percolations in cluster merging process~\cite{RER_Park, BFW_Choi}. On the other hand, the hybrid transition at the CE line can be found in the epidemic model~\cite{ED_Choi}. Thus, our results integrate such scattered previous results.  

The two characteristic transition lines are merged at $x_{\rm TP}$ and $q_c=2$. We revealed that the two transition lines exhibit a singular behavior $x_h-x_\ell(q) \sim (q_c-q)^{2/3}$. In percolation limit ($q\to 1^+$), the discontinuous transition for $x > x_{1\ell} = e^2$ is induced by the bi-stability of the hidden spin states, whereas in Ising limit ($q\to 2$), the discontinuous transition for $x_\ell < x < e^2$ is induced by the competition between the interaction and entropy of the system. 

This paper highlights how introducing hidden states to the Potts model can lead to a more versatile and powerful tool for modeling a wide range of complex phenomena in theoretical and practical contexts, including those related to non-equilibrium systems and political elections. This research could significantly improve our understanding of real-world systems and make more accurate predictions or decisions in various fields.

\begin{acknowledgments}
We would like to thank Prof. J\'anos Kert\'esz for introducing this problem and Prof. Maxi San Miguel for discussing voter models. This work was supported by the National Research Foundation of Korea by Grant No. NRF-2014R1A3A2-069005 and RS-2023-00279802 and the KENTECH Research Grant No. KRG-2021-01-007 (BK), and a KIAS Individual Grant(No. CG079902) from Korea Institute for Advanced Study (DSL).
\end{acknowledgments}



\newpage

\onecolumngrid

\SupplementaryMaterials

\setcounter{equation}{0}
\setcounter{figure}{0}
\setcounter{table}{0}
\setcounter{page}{1}
\makeatletter
\renewcommand{\theequation}{S\arabic{equation}}
\renewcommand{\thefigure}{S\arabic{figure}}

\begin{center}
\textbf{\large Supplemental Information for \\ Entropy-Induced Phase Transitions in a Hidden Potts Model}
\end{center}

\section{Coefficients of $B_n$ and $C_n$ in the free energy expansion}
\label{seca:Cn}

We first expand $m_r(m)$, the solution to \eqref{eq:equationOfState_R_Hidden}, for small $m$ as $m_r(m)=m_r(0)+\sum_{n=2}^\infty B_n \left(\dfrac{q m}{\tau}\right)^n$ with the coefficients given by 
\begin{align}
B_2 &= -{(q-1) m_r(0) (1-m_r(0)) \over 2q^2 \{1 - {1\over\tau} m_r(0) \left(1-m_r(0)\right)\}}, \cr
B_3 &= - {(q-1)(q-2) m_r(0) (1-m_r(0)) \over 6q^3 \{1 - {1\over\tau} m_r(0) \left(1-m_r(0)\right)\}},\cr
B_4 &= - {(q-1) m_r(0) (1-m_r(0)) \over 24 q^4 \{1 - {1\over\tau} m_r(0) \left(1-m_r(0)\right)\}}\left[q^2 - 6q + 6 - {3 (q-1) (1 - 2m_r(0)) \over \{1 - {1\over\tau} m_r(0) \left(1-m_r(0)\right)\}^2}\right],\cr
B_5 &= -{(q-1)(q-2) m_r(0) (1-m_r(0)) \over 120 q^5 \{1 - {1\over\tau} m_r(0) \left(1-m_r(0)\right)\}} \left[q^2 -12q +12-{10 (q-1) (1 - 2m_r(0)) \over \{1 - {1\over\tau} m_r(0) \left(1-m_r(0)\right)\}^2}\right],\cr
B_6 &= -{(q-1) m_r(0) (1-m_r(0)) \over 720 q^6 \{1 - {1\over\tau} m_r(0) \left(1-m_r(0)\right)\}}\left[ q^4 - 30 q^3 + 150 q^2 - 240 q + 120 +{45 (1-q)^2 (1 - 4\tau) \over \{1 - {1\over\tau} m_r(0) \left(1-m_r(0)\right)\}^4}\right. \cr
& \hspace{75pt} \left.- {30 (1-q)^2 (1 - 9\tau) \over \{1 - {1\over\tau} m_r(0) \left(1-m_r(0)\right)\}^3} - {5(q-1) \{18 (q-1) \tau  + (1-2m_r(0)) (5q^2 -26 q + 26) \} \over \{1 - {1\over\tau} m_r(0) \left(1-m_r(0)\right)\}^2}\right].
\end{align}
Here $m_r(0)$ can be obtained by solving \eqref{eq:equationOfState_R_Hidden} with $m=0$. Inserting this expansion into \eqref{eq:equationOfState_m_Hidden}, one can obtain \eqref{eq:free_energy_density} with the coefficients given by 
\begin{align}
C_0 &= {1\over 2\tau} (1-m_r(0))^2 - \ln\left({r\over m_r(0)} \right),\cr
C_1 &=0,\cr
C_2 &= {q-1\over 2q^2}\left\{{\tau} - 1+m_r(0)\right\}  ,\cr
C_3 &= -{(q-1)(q-2) (1-m_r(0))\over 6 q^3} ,\cr
C_4 &= -{(q-1)(1-m_r(0)) \over 24 q^4} \left[q^2-6q+6 + {3 (q-1) m_r(0) \over 1 - {1\over\tau} m_r(0) \left(1-m_r(0)\right)}\right],\cr
C_5 &= -{(q-1)(q-2) (1-m_r(0))\over 120 q^5} \left[q^2 -12 q + 12 +{10 (q-1) m_r(0)\over 1 - {1\over\tau} m_r(0) \left(1-m_r(0)\right)}\right],\cr
C_6 &= -{(q-1) (1-m_r(0)) \over 720 q^6 } \left[q^4 - 30q^3 + 150 q^2 -240 q + 120-15{\tau} { (1-q)^2 (2 - m_r(0)/\tau) \over \{1 - {1\over\tau} m_r(0) \left(1-m_r(0)\right)\}^3 } \right.\cr
& \left. \hspace{150pt} + 30 {\tau}{(1-q)^2 \over\{1 - {1\over\tau} m_r(0) \left(1-m_r(0)\right)\}^2}+ 5{(q-1)(5q^2 -26 q + 26)m_r(0)\over 1 - {1\over\tau} m_r(0) \left(1-m_r(0)\right)}\right].
\label{eq:Cn}
\end{align}

\section{Derivation of $\beta^{\prime}$ at CP}
\label{seca:CP_Crit}

The order parameter $m^*(\tau)$ is the value of $m$ where the free energy $f(m,\tau)$ is minimized for given temperature $\tau$, as given in \eqref{eq:argmin}, satisfying ${\partial f\over \partial m}|_{m^*}=0$. Its derivative ${dm^*\over d\tau}$ with respect to temperature $\tau$ can be represented in terms of the second-order derivative of $f$ as
\begin{align}
{d m^*\over d\tau} =  - {{\partial^2 f \over \partial m \partial \tau} \over {\partial^2 f\over \partial m^2}}\bigg|_{m^*(\tau)},
\end{align} 
because
\begin{align}
{d\over d\tau} \left({\partial f\over \partial m}\bigg|_{m^*(\tau)}\right) = {\partial^2 f \over \partial m^2}\bigg|_{m^*} {dm^*\over d\tau} + {\partial^2 f \over \partial m \partial \tau}\bigg|_{m^*}=0.
\end{align}

At CP ($x=x_\ell$ and $\tau = \tau_{c}^{\prime}$), the derivative ${dm^*\over d\tau}$ diverges; i.e. $dm^*/d\tau \to \infty$, and so both the first- and second-order derivative of the free energy $f(m,\tau)$ to $m$ are zero there, i.e., ${\partial f \over \partial m}\bigg|_{\textrm{CP}}={\partial f^2 \over \partial m^2}\bigg|_{\textrm{CP}}=0$. Let us denote the order parameter at CP by $m_c^\prime$.  Near CP, the free energy $f$ will be expanded around $m_c^\prime$ and $\tau_c^\prime$ as
\begin{align}
f(m &= m_c^\prime+\delta m, \tau = \tau_c^\prime+\delta \tau) 
= f(m_{c}^{\prime}, \tau_{c}^{\prime}) + \dfrac{\partial f}{\partial \tau} \bigg|_{\textrm{CP}} \delta \tau + \dfrac{\partial^2 f}{\partial m \partial \tau} \bigg|_{\textrm{CP}} \delta \tau \delta m + \dfrac{\partial^3 f}{\partial m^3} \bigg|_{\textrm{CP}} \dfrac{1}{6}\delta m^3+ \textrm{ H.O.},
\end{align}
with $\delta \tau = \tau - \tau_{c}^{\prime}$ and $\delta m = m - m_{c}^{\prime}$. 
Differentiating with respect to $\delta m$, we obtain 
\begin{align}
\dfrac{\partial f}{\partial m} 
\simeq \dfrac{\partial^2 f}{\partial m \partial \tau} \bigg|_\textrm{CP} \delta \tau + \dfrac{\partial^3 f}{\partial m^3} \bigg|_\textrm{CP} \dfrac{1}{2}\delta m^2+ \textrm{ H.O.} 
\end{align}
which can be zero at $m^*$ if $\delta \tau \sim - \delta m^2$. Therefore, one can see that $m^*-m_c^\prime\sim (\tau_c^\prime - \tau)^{1/2}$, which indicates  $\beta^{\prime} = 1/2$ at CP in \eqref{eq:CP_Criticality}. 

\begin{figure*}
\centering
\resizebox{1.0\columnwidth}{!}{\includegraphics{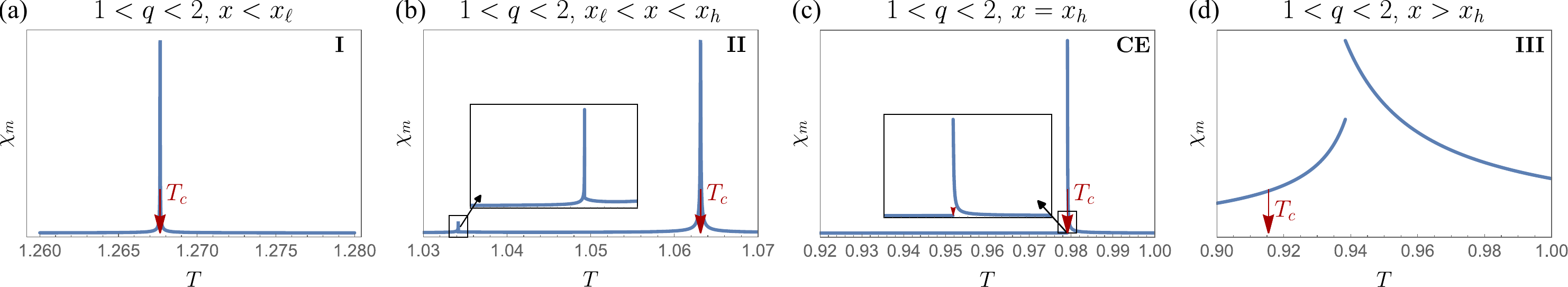}}
\caption{  
Plots of the susceptibility $\chi_m$ vs $T(=z\tau/q)$ for $z=4$ and $1<q=1.5<2$. 
(a) When $x=3.0 < x_\ell$, $\chi_m$ diverges; (b) when $x=x_\ell\approx 4.1$, $\chi_m$ as evidence of the second-order transition, this is a figure showing how the susceptibility $\chi_{m}$ changes with temperature $T$ according to various $r$ values. 
(a) For $x=3.0$, $\chi_{m}$ diverges to $\infty$ at $T_{c\pm}$.
(b) For $r=4.1$, $\chi_{m}$ diverges to $\infty$ at $T_{c\pm}$ and also shows discontinuous jump at $T_{f}$ in which $m$ so does.
(c) For $x=4.685$, $\chi_{m}$ shows discontinuous jump at $T \rightarrow T_{c-}$ ($m$ abruptly jumps to $0$) and divergence at $T \rightarrow T_{c+}$. This is why we call this type of PT the HPT.
(d) For $x=5.2$, $\chi_{m}$ shows discontinuous jump at $T_{f} (> T_{c})$.
\label{fig:fig5}	
} 
\end{figure*}

\section{Susceptibility}
\label{seca:chi}

For a non-zero external field,  the order parameter $m^*(h)$ can be determined by searching for the value of $m$ minimizing the free energy $f(m, h) \equiv f(m, \tilde{m}_r(m),h)$ with $\tilde{m}_r(m)$  the solution to \eqref{eq:equationOfState_R_Hidden} for given $m$ and $h$. With $f(m,h)$, one can characterize
the  response of the order parameter (magnetization) to an external field (magnetic field) by the susceptibility defined as
\begin{align}
\chi_m = {\partial m^*(h)\over \partial h}\bigg|_{h=0}.
\end{align}

From \eqref{eq:equationOfState_R_Hidden} 
one can obtain 
$\tilde{m}_{r}(m,h) = \tilde{m}_{r}(0,h) + \sum_{n=1}^\infty \tilde{B}_n(h) \left(\dfrac{q m}{\tau}\right)^n$ where $\tilde{m}_{r}(0,h)$ is given as
\begin{align}
\tilde{m}_{r}(0,h) &= m_r(0) + {h\over q} \dfrac{m_r(0) (1-m_r(0))}{1 - {1\over\tau} m_r(0) \left(1-m_r(0)\right)} + O(h^2)
\end{align}
up to the linear order of $h$ and $\tilde{B}_n(h) = B_n + O(h)$. Inserting $\tilde{m}_{r}(m,h)$ into \eqref{eq:fmR}, we obtain
\begin{align}
f(m,H) = f(m,m_{r}(m,h),h) = \sum_{n=0}^\infty \tilde{C}_n(h) \left(\dfrac{q m}{\tau}\right)^n
\end{align}
with the coefficients given by 
\begin{align}
\tilde{C}_0 &= C_0  
\cr
\tilde{C}_1 &= - {h\over q^2} (q-1)(1-m_r(0)) + O(h^2)\cr
\tilde{C}_2 &= C_2 - {h\over 2q^3} (q-1) (q-2) (1-m_r(0)) + O(h^2)\cr
\tilde{C}_3 &= C_3 - {h\over 6q^4} (q-1) (1-m_r(0)) \left[q^2 - 6q + 6 + {3  (q-1) m_r(0)\over 1 - {1\over\tau} m_r(0) \left(1-m_r(0)\right)}\right] + O(h^2)
\label{eq:tildeCn}
\end{align}

Since $f(m, h)$ is minimized at $m^{*}(h)$, it holds that $\partial f/\partial m=0$ at $m^{*}(h)$. Considering that ${d\over dh}\left({\partial f\over \partial m}|_{m^*(h)}\right)=f_{mm} {dm^*\over dh} + f_{mh} =0$ and using the expansion of the free energy in the above, we can represent the susceptibility near $\tau_c$ for $1<q<2$ as follows: 
\begin{align}
\chi_m 
&= -\dfrac{f_{mh}}{f_{mm}}\bigg|_{m=m^{*}(0), h=0}
\simeq {\tau }{(q-1)(1-m_r(0)) \over q^3\left\{2C_2 + 6 C_3 (m/ \tau)\right\}}, 
\end{align}
where
\begin{align}
f_{mh} \equiv \frac{\partial^2 f(m,h)}{\partial m \partial h} \quad {\textrm{and}} \quad f_{mm} \equiv \frac{\partial^2 f(m,h)}{\partial m^2}.
\end{align}
Using \eqref{eq:CPTsmallr}, we obtain that for small $x$,
\begin{align}
\chi_m&\simeq \begin{cases}
\dfrac{\left(1-m_r(0)\right)\tau}{q^2\{1 - (1-m_r(0))/\tau\}} \sim \dfrac{1}{\tau-\tau_c} & \ {\rm for} \ \tau>\tau_c\\ \\
\dfrac{-\left(1-m_r(0)\right)\tau}{q^2\{1 - (1-m_r(0))/\tau\}} \sim \dfrac{1}{\tau_c-\tau} & \ {\rm for} \ \tau<\tau_c,
\end{cases}
\end{align}
for the continuous transition. 
At the CE point ($x=x_h$ and $\tau=\tau_c$), $m$ shows a discontinuous jump at $\tau_c^-$, and therefore, the susceptibility behaves as 
\begin{align}
\chi_m&\simeq \begin{cases}
\dfrac{\left(1-m_r(0)\right)\tau}{q^2\{1 - (1-m_r(0))/\tau\}} \sim \dfrac{1}{\tau-\tau_c} & \ {\rm for} \ \tau>\tau_c\\ \\
{\rm constant}  & \ {\rm for} \ \tau<\tau_c
\end{cases}
\end{align}
as shown in Fig.~\ref{fig:fig4}(c). Accordingly, the critical exponent $\gamma_m^{+}=1$. Then, the scaling relation $\alpha+2\beta_m+\gamma_m^+=2$ holds (see Table 1). 

\begin{table}[h]
\centering
\begin{normalsize}
\setlength{\tabcolsep}{5pt}
{\renewcommand{\arraystretch}{1.8}
\begin{tabular}{@{\extracolsep{\fill}}cccccccc} 
\hline
\hline
Range of $q$&\multicolumn{1}{c}{$\alpha_m$}&\multicolumn{1}{c}{$\alpha_{m_{r}}$}&\multicolumn{1}{c}{$\beta_m$}&\multicolumn{1}{c}{$\beta_{m_{r}}$}&
\multicolumn{1}{c}{$\gamma_{m\pm}$}& \multicolumn{1}{c}{$\gamma_{{m_{r}}\pm}$} \cr
\hline
\hline
$1 \le q < 2$&$-1$&$-1$&$1$&$2$&$1$&$0$ \cr\
$q = 2, r<r_{c}$&$0$&$0$&$\frac{1}{2}$&$1$&$1$&$\frac{1}{2}$\cr\
$q = 2, r = r_{c}$&$\frac{1}{2}$&$\frac{1}{2}$&$\frac{1}{4}$&$\frac{1}{2}$&$1$&$\frac{3}{4}$\cr
\hline
\hline
\end{tabular}}
\end{normalsize}
\caption{
Critical exponents for various types of CPT: Here, $\alpha$ is the exponent of the specific heat, $\beta_m$ ($\beta_R$) is the exponent of the magnetization $m$ ($R$) at zero external magnetic fields, and $\gamma_{m}$ ($\gamma_{R}$) is the exponent of the susceptibility for $m$ ($R$)-magnetization near the transition temperature. \label{ta:exponents}}
\end{table}

\section{Phase transitions and supercritical behavior near TP}
\label{seca:TP}
In this section, we study the phase transitions at and near the TP considered in Sec.~\ref{subsec:q2} shown in Fig.~\ref{fig:fig1}(b). Let us first consider the case of $q=2$. Then $C_3=C_5=0$ from \eqref{eq:Cn} and one can represent the free energy as  
\begin{align}
f(m) \simeq C_2 \left(\dfrac{qm}{\tau}\right)^2 + C_4 \left(\dfrac{qm}{\tau}\right)^4 + C_6 \left(\dfrac{qm}{\tau}\right)^6.
\end{align}
The two coefficients $C_2$ and $C_4$ become zero when $x=x_{\rm TP}$ and $\tau=\tau_{\rm TP}$  with 
\begin{align}
x_{\rm TP}={2e\over 3}, \, \tau_{\rm TP}={3\over 5} \ {\rm and } \ m_{r}(0,\tau_{\rm TP}).
\end{align}
At $x_{\rm TP}$, using \eqref{eq:equationOfState_R_Hidden}, both $\epsilon_\tau \equiv \tau - \tau_{\rm TP}$ and $\epsilon_r \equiv m_{r}(0,\tau) - m_r(0,\tau_{\rm TP})$ have the same order of magnitude, and thus within the first order of $\epsilon_\tau$, we obtain the following:
\begin{align}
C_2 \propto \epsilon_\tau,\ \, \ C_4 \propto \epsilon_\tau, \ \, \ \textrm{ and } \ \, \ C_6 \propto \mathcal{O}(1).
\end{align} 
When $\tau \to \tau_{\rm TP}$, because the order of two coefficients $C_2$ and $C_4$ are the same, the fourth-order term can be ignored compared with $C_2$. When $C_2>0$, the global minimum locates at $m=0$, while $C_2<0$, and $C_6>0$ generate the global minimum at finite $m$ as 
\begin{align}
m\sim \begin{cases}
0 & \ {\rm for} \quad \tau > \tau_{\rm TP},\\
(-\epsilon_\tau)^{1/4} & \ {\rm for} \quad  \tau<\tau_{\rm TP}. 
\end{cases}
\end{align}
This result is equivalent to \eqref{eq:CPTatTP}.

As $q$ and $x$ deviate slightly from $q_c=2$ and $x_{\rm TP}$ along with the CP and CE lines, $C_3$ and $C_4$ terms increase gradually and contribute to the local minimum of $f(m)$, together with $C_2$ and $C_6$ terms. This implies that the four terms would have the same order of magnitudes. Moreover, using $C_6\sim \mathcal{O}(1)$, we presume that $C_2 = \mathcal{O}(m^4)$, $C_3 = \mathcal{O}(m^3)$, and $C_4 = \mathcal{O}(m^2)$. 
From \eqref{eq:Cn}, $C_4 \sim \epsilon_\tau$ and $C_3 \sim (2-q) = \epsilon_q$, and thus $\epsilon_\tau \sim \mathcal{O}(m^2)$ and $\epsilon_q \sim \mathcal{O}(m^3)$. On the other hand,  
$C_2 \approx \epsilon_\tau+\epsilon_r$, and hence, $\epsilon_r = - \epsilon_\tau + \mathcal{O}(m^4)$.
Using \eqref{eq:equationOfState_R_Hidden}, $\mathcal{O}(\epsilon_x) = \mathcal{O}(\epsilon_\tau) = \mathcal{O}(m^2)$. Therefore, we obtain that $\mathcal{O}(\epsilon_x) = \mathcal{O}(\epsilon_q^{2/3})$ on the CP and CE lines.  
\begin{align}
x_{\ell} - x_{\rm TP} \sim (q_c - q)^{2/3} \ {\rm and } \ x_{h} - x_{\rm TP}  \sim (q_c - q)^{2/3}.
\end{align}

\section{Phase transitions of the order parameter in hidden states \\ in the $q\to1^{+}$ limit}
\label{seca:OHS}
To obtain  $m_{r}$ in the $q \to 1^{+}$ limit, we consider the constrained free energy in \eqref{eq:fmR} with $m$ set to zero, 
\begin{align}
f(0,m_r) \equiv f(m_r) = {m_r^2-1\over 2\tau} - \ln (q + r e^{-{1-m_r\over \tau}}),
\end{align}
and the global minimum location  
\begin{align}
\bar{m}_r = \arg \min_{m_r} f(m_r).
\end{align}
It should satisfies $f_{m_r}(m_r)|_{m_r=\bar{m}_r}=0$, which is equivalent to \eqref{eq:equationOfState_R_Hidden} with $m=0$ or
 \begin{align}
\bar{m}_r={xe^{-(1-\bar{m}_r)/\tau}\over 1+ x e^{-{(1-\bar{m}_r)/\tau}}}.
\label{eqa:F}
\end{align}
Also, the global minimum condition requires 
\begin{align}
f_{m_r m_r}(m_r)\bigg|_{m_r=\bar{m}_r} = {1\over \tau} \left[1 - {\bar{m}_r(1-\bar{m}_r)\over \tau} \right]>0.
\end{align} 

The solution $\bar{m}_r$ of \eqref{eq:F} implies the following: ($x=0$, $\bar{m}_r=0$) is the solution of the original Q-Potts model representing the disordered (or paramagnetic) phase. For the QR-Potts model. 

\begin{figure*}
\centering
\resizebox{1.0\columnwidth}{!}{\includegraphics{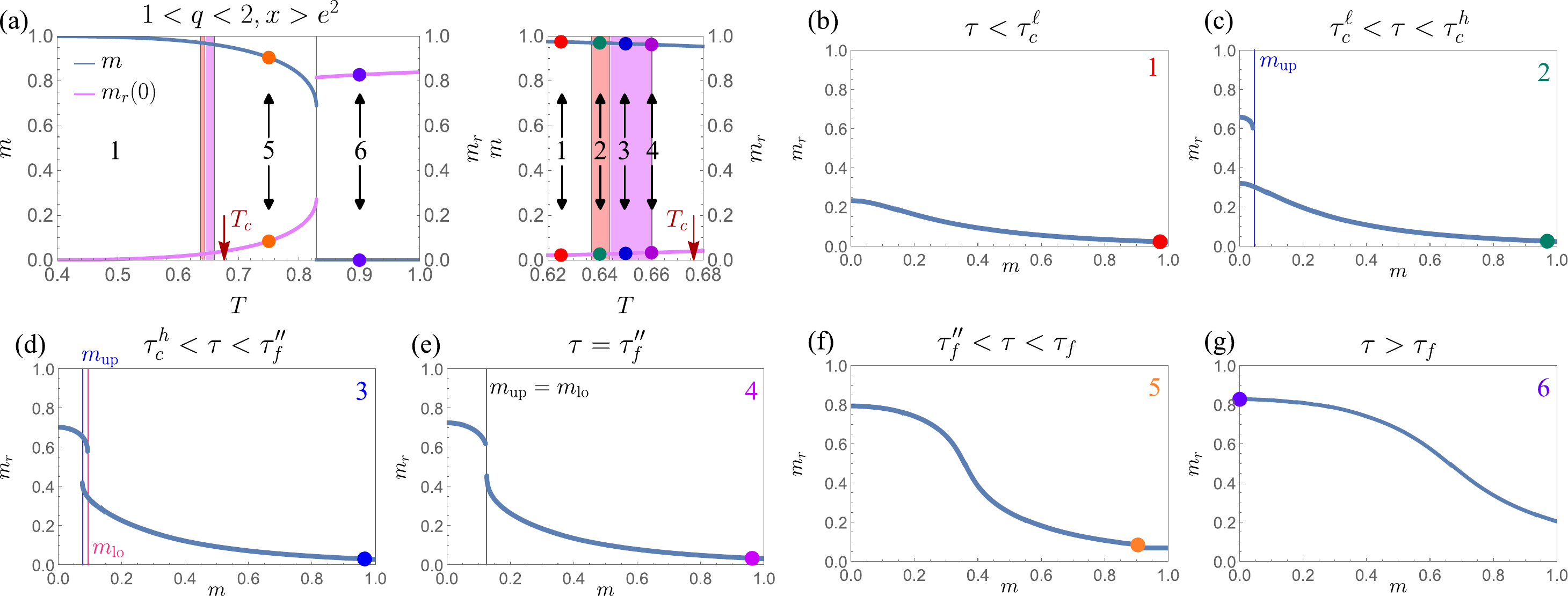}}
\caption{  
(a) Plots of the order parameters $m$ (left scale) and $m_r(0)$ (right scale) vs $T= z \tau / q$ for $1 < q < 2$ and $x > e^2$. The order parameter $m$ ($m_r(0)=0$) decreases (increases) continuously from $m=1$ ($m_r(0)$) as $T$ is increased. At $T_f$, they exhibit a discontinuous drop to zero (jump) and then stay at zero (continuously increases). 
(b)--(g) Plot of the order parameter $m_r(m)$ that the system is in hidden states vs $m$ for $q=1.5$ and $x=8>e^2$. 
(b) For $\tau=0.62 < \tau_{c}^{\ell}$, only one $m_{r}(m)$ (lower) curve appears in system.
(c) For $\tau_{c}^{\ell} < \tau=0.64 < \tau_{c}^{h}$, another $m_{r}(m)$ curve emerges above the lower $m_{r}(m)$ curve in $m \in (0,m_{\up})$.
(d) For $\tau_{c}^{h} < \tau=0.65 < \tau_{f}^{\prime\prime}$, the lower $m_{r}(m)$ curve separates from $m=0$ line ($y$-axis), and so exists only in $m \in (m_\lo, 1)$.
(e) For $\tau=0.66 = \tau_{f}^{\prime\prime}$, two disconnected $m_{r}(m)$ curves begin to merge into one curve.
(f) For $\tau_{f}^{\prime\prime} < \tau=0.75 < \tau_{f}$, only one $m_{r}(m)$ curve appears in the system. However, the value of $m_{r}(m)$ is much larger than that of case (b).
(g) For $\tau=0.9 > \tau_{f}$, a position of $m$ and $m_{r}$ on $m_{r}(m)$ curve jump from an ordered state to a disordered state.
\label{fig:fig6}	
} 
\end{figure*}

(i) When $0< x \ll 1$, $xe^{(1-m_r)/\tau}$ is small in the entire range of $m_r$ and $\tau$. Therefore, $m_r$ is also always small compared with $\mathcal{O}(1)$. 
The free energy exhibits a single global minimum, and only one stable solution $\bar{m}_r$ appears near zero as $\bar{m}_r \approx xe^{-(1-\bar{m}_r)/\tau} \approx xe^{-1/\tau}$. 
As shown in Fig.~\ref{fig:fig6}(a), $m_r$ increases gradually with $\tau$ increasing. 

(ii) As $x$ is increased further and arrives at $e^2$, there exists $\bar{m}_r$ such that 
\begin{align}
    f_{m_r}(m_r)|_{m_r=\bar{m}_r}=f_{m_r m_r}(m_r)|_{m_r=\bar{m}_r}=0. 
\end{align}
This is a transition point corresponding to CP. Similar to $q\to2^{-}$ case, this critical point can be obtained from the condition $f_{m_r m_r}(m_r)|_{\tau=\tau_c^{\prime}}=0$. 
Therefore, $\tau_c^{\prime}$ is obtained as follows:  
\begin{align}
\tau_c^{\prime}=\bar{m}_r(x,\tau_c^\prime)(1-\bar{m}_r(x,\tau_c^{\prime})) = \dfrac{1}{4}.
\label{eq:tauc}
\end{align}
At $\tau_c^{\prime}$, $m_r$ increases most rapidly as shown in Fig.~\ref{fig:fig6}(b), because $dm_r/d\tau \propto (f_{m_r m_r}(m_r))^{-1} \to \infty$ diverges at $\tau_c^{\prime}$. 
As a result, $x_{1\ell}$ ($x_\ell$ for $q=1$) is $e^2$ as shown in Fig.~\ref{fig:fig1}(b).

(iii) When $x > e^2$, there exist two solutions $\bar{m}_{r-}$ and $\bar{m}_{r+}$ which satisfy the conditions $f_{m_r}(m_r)=0$ and $f_{m_r m_r}(m_r) > 0$. When $\tau$ is relative low (high), $f(\bar{m}_{r-})$ is smaller (larger) than $f(\bar{m}_{r+})$. So the global minimum occurs at $\bar{m}_{r-}$ and $\bar{m}_{r+}$ for lower and higher $\tau$, respectively. 
As the position of the global minimum changes from $\bar{m}_{r-}$ to $\bar{m}_{r+}$ at $\tau_f$, the order parameter $\bar{m}_r$ shows a discontinuous jump, and thus the first-order transition occurs. 

We consider several extreme cases for which $m_r$ is analytically obtained. In case (i), when $x \equiv r/q \ll 1$, using Eq.~(\ref{eq:F}), we obtain 
\begin{align}
m_r \simeq xe^{-1/\tau}
\end{align} 
for all $\tau$. In case (iii), when $x > e^2$, 
Eq.~(\ref{eq:F}) has two solutions $\bar{m}_{r-}$ and $\bar{m}_{r+}$ when $\tau$ is between two characteristic $\tau_-$ and $\tau_+$. (see Fig.~\ref{fig:fig6}(c)). At $\tau_-$, $\bar{m}_{r+}$ appears as a local (not global yet) minimum of $f(m_r)$. Thus, 
\begin{align}
&f_{m_r}(m_r)|_{\bar{m}_{r+}}=f_{m_r m_r}(m_r)|_{\bar{m}_{r+}}=0 \cr
&\tau_-= \bar{m}_{r+}(1-\bar{m}_{r+}), \text{ with }
\bar{m}_{r+} =\dfrac{xe^{-{1/\bar{m}_{r+}}}}{1 + xe^{-{1/\bar{m}_{r+}}}}.
\label{eq:TellR*}
\end{align}
At $\tau_+$, the smaller solution $\bar{m}_{r-}$ disappears, and thus the larger solution $\bar{m}_{r-}$ becomes a unique local minimum of $f(m_r)$. Thus,  
\begin{align}
f_{m_r}(m_r) |_{\bar{m}_{r-}}=f_{m_r m_r}(m_r)|_{\bar{m}_{r-}}=0
\end{align}
Using this relation, we can also find that
\begin{align}
\tau_+ = \bar{m}_{r-}(1-\bar{m}_{r-}), \ {\rm with} \ 
\bar{m}_{r-} = \dfrac{xe^{-{1/\bar{m}_{r-}}}}{1 + xe^{-{1/\bar{m}_{r-}}}}.
\label{eq:ThR*}
\end{align}
From Eqs.~(\ref{eq:TellR*}) and~(\ref{eq:ThR*}), we find that $\bar{m}_{r-}$ and $\bar{m}_{r+}$ are obtained as follows:
\begin{align}
\bar{m}_{r-}\simeq  {1\over \ln \left( {x} \ln x \right)} \ {\rm and}
\quad
\bar{m}_{r+} \simeq 1 - \dfrac{e}{x} 
\label{eq:R*}
\end{align}
and we can obtain \eqref{eq:mr0}.

\section{Bi-stability of the order parameter in hidden states with large $x > e^2$}
\label{seca:V_HS}

We investigate the behavior of $m_{r}(m)$ as a function of $\tau$, but $x$ is fixed in the regime $x >e^2$. As one compares Fig.~\ref{fig:fig6}(b) with (c), a new $m_{r}(m)$ curve appears at $m=0$ as $\tau \to \tau_c^{\ell}$. This emergence implies that a discontinuous transition of $m_r(m)$ occurs at $m=0$ as $\tau$ increases. As $\tau$ increases beyond $\tau_c^{\ell}$, the upper curve of $m_r(m)$ is somewhat extended to $m_{\up}$ as shown in Fig.~\ref{fig:fig6}(c). As $\tau \to \tau_c^{h}$, the lower curve is detached from a vertical line of $m=0$, and so, the lower curve spans from $m_{\lo}$ to $1$. Note that $m_{\lo}$ is less than $m_{\up}$ for $\tau_c^{h} < \tau < \tau_{f}^{\prime\prime}$ as shown in Fig.~\ref{fig:fig6}(d). At $\tau_{f}^{\prime\prime}$,  $m_{\up}$ and $m_{\lo}$ meet as shown in Fig.~\ref{fig:fig6}(e).
Finally, when $\tau$ is higher than $\tau_{f}^{\prime\prime}$, two curves are merged into one curve as shown in Fig.~\ref{fig:fig6}(f), and thus, the local minimum with a high degree of $m_{r}$ state disappears.
From this behavior of $m_{r}$ curves, we find that the meta-stable state sustains in a range of $\tau_c^{\ell} < \tau < \tau_{f}^{\prime\prime}$ only for $x>e^2$.

\newpage

\twocolumngrid

\end{document}